# Using 3D Printing to Visualize Social Media Big Data

Zachary J. Weber and Vijay Gadepally
MIT Lincoln Laboratory
Email Addresses: {zweber, vijayg}@ll.mit.edu

*Abstract*—Big data volume continues to grow at unprecedented rates. One of the key features that makes big data valuable is the promise to find unknown patterns or correlations that may be able to improve the quality of processes or systems. Unfortunately, with the exponential growth in data, users often have difficulty in visualizing the often-unstructured, non-homogeneous data coming from a variety of sources. The recent growth in popularity of 3D printing has ushered in a revolutionary way to interact with big data. Using a 3D printed mockup up a physical or notional environment, one can display data on the mockup to show real-time data patterns. In this poster and demonstration, we describe the process of 3D printing and demonstrate an application of displaying Twitter data on a 3D mockup of the Massachusetts Institute of Technology (MIT) campus, known as LuminoCity.

## I. INTRODUCTION

Advances in sensor technology have allowed the rapid expansion of information collected from a variety of sources. These sensors are fueling the big data revolution, which is characterized by the exponential growth of volume, velocity and variety. These big data sets are valuable because of the promise to provide deep insight into important patterns, correlations or aberrations that may be present in a system. In order to extract valuable information, a user must ask the correct questions. This is often troublesome because users do not know what to ask of the data.

Information visualization is the study of using visual tools to reinforce the ability of humans to extract meaningful information from potentially abstract sources. In [1], the authors provide a survey of techniques and recent advances in the field of information visualization. One of the new tools for rapid prototyping – 3D printing [2] – can be used as a means to visualize data that can be represented by 3-Dimensional (3D) surfaces. For example, in [3], the authors describe the usage of 3D scanning to model building data. In [4], the authors show an application in which 3D printing is used to model a heart derived from patient data. The authors of [5] describe the use of digital holography for visualizing 3D information. The properties of 3D printing make it particularly useful in modeling real world physical entities such as buildings, objects, cities, etc.

Fascination with 3D models of cities predates the first human aerial ascent by centuries. Today, cityscapes of monumental extent capture the limelight. For example, the 1964 Worlds-Fair-commissioned, 1:1200 scale Panorama of the City of the New York [6], now at the Queens Museum, comprises 830,000 buildings, with a model size exceeding 1000 m$^2$. Although static models have predominated, active lighting has recently been incorporated in 3D model cities. The London Building Centre's city model, developed for the 2012 Olympics, employs changing overhead spot lighting, and the Shanghai Urban Planning Exhibition Center's future city model utilizes individual LEDs embedded within the scale buildings. Recently, modelers have begun leveraging 3D printing to generate individual model buildings. In [7], a team of artists modeled 1000 downtown Chicago buildings to recreate a faithful model of the Windy City. In [8], we describe LuminoCity, a novel three-dimensional data display created by 3D printing LADAR data of the MIT campus (Figure 1).

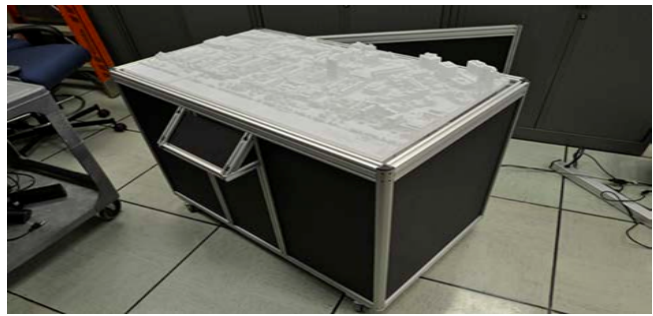

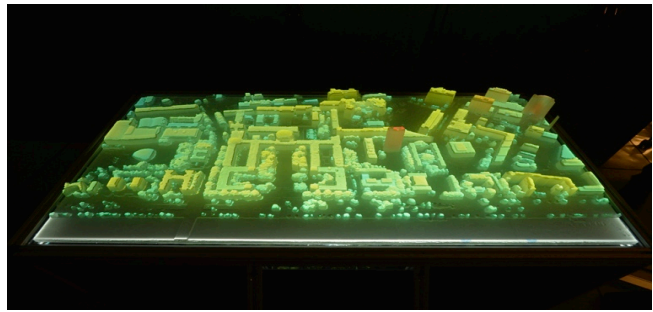

Fig. 1. (a) LuminoCity with Projector Base (b) LuminoCity illuminated by measured height above ground level (data generated by LADAR)

## II. POSTER AND DEMONSTRATION

LuminoCity is a platform for displaying a wide variety of data and analysis. In this sense, it combines the flexibility and adaptability of conventional 2D display screens with the immersive experience of static 3D models. While 3D glasses and other technologies allow a single user to view a scene with the appearance of three dimensions, they are not suitable for collaborations involving multiple users. In contrast, LuminoCity provides all members of a team with a shared, tangible 3D visualization. As an example of LuminoCity's potential applications, imagine a scenario where a city's traffic manager needs to work with the city's urban planner and representatives from utility companies in order to investigate traffic patterns in a crowded neighborhood derived from a variety of sensors such as traffic lights, social media, street cameras, mobile phones, etc. The ability to project these varied data sources onto a 3D representation of the area in question can rapidly aid in gaining insight that can be used to alleviate traffic hot spots.

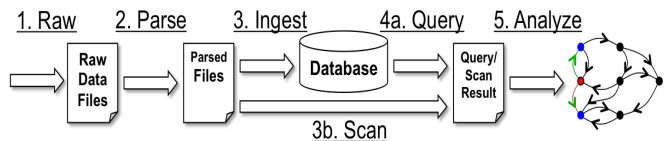

Fig. 2. General purpose big data processing pipeline.

In this demonstration, we use the LuminoCity 3D display to visualize data analysis derived from the Twitter Decahose. Twitter is a popular microblogging website, and the Twitter Decahose [9] provides 10% of all tweets, chosen at random, for paying customers. Using Twitter from a mobile device allows one to share their location in addition to the tweet (user's post). Purchasing the Twitter Decahose also allows one to collect targeted datasets that satisfy certain criteria such as the inclusion of location information. The use of location in conjunction with content is a very active area of research and we target all geo-located tweets. Raw data received from the Twitter Decahose is then parsed into a tab separated value (TSV) format and further parsed into a format that can be used to store these tweets on a local filesystem or insert them into a high performance database (DB) such as Apache Accumulo [10]. These steps of the data processing pipeline are shown in Figure 2 as steps 1, 2 and 3. After data is stored in a database, one can query data that satisfies certain criteria such as location.

Tweets that are geo-tagged inside the geographic area covered by LuminoCity are selected from the stored and parsed data using the Dynamic Distributed Dimensional Data Model (D4M) [11]. D4M is a high performance schema and software package that interfaces GNU Octave or MATLAB with the Apache Accumulo database. The D4M syntax allows for easy data filtering by latitude and longitude. Data is stored in database tables (similar to a large spreadsheet) and are indexed by the DB for efficient query performance. For example, with a database table T, and latitude and longitude ranging from (+42.350, -71.090) to (+42.357, -71.099) which roughly corresponds to the MIT campus, the D4M database query will be:

```
range=latlon|+-4721..305900:+-4721..305979;
Amit = T(:,range);
```

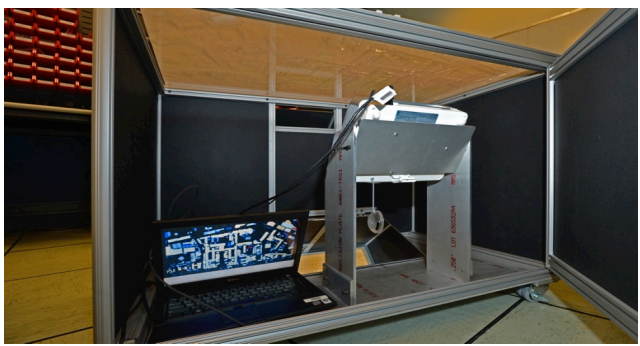

Fig. 3. Base of LuminoCity 3D model of MIT campus. View of prototype interior for LuminoCity showing the optics and computer driven projector.

Where `Amit`, is a data structure that contains all tweets within the lat-lon box specified by the variable `range`. Once the database query is performed, the precise latitude and longitude of each tweet is extracted from the filtered tweets along with user text. This information is then fed into a program that aligns the tweets with the LuminoCity grid for display. This aligned data is then loaded onto a laptop computer housed in the base of LuminoCity (Figure 2) that drives a high power projector that lights up elements based on feedback generated from a mounted tablet as shown in Figure 3.

For example, this data set will be mined to display data that may be important for administrators at MIT who wish to know about patterns on campus through key-word searches and topic clustering. Other demonstrations may include animating twitter traffic volume as a function of time and space to provide insight into campus patterns or life.

LuminoCity in its full form is shown in Figure 4 with projection device enabled. Users can interact with LuminoCity via an attached touch screen control as shown in Figure 4.

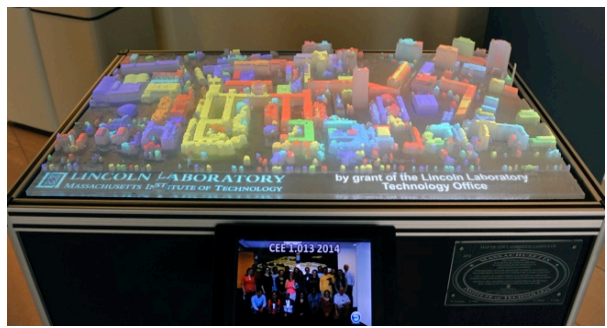

Fig. 4. Interactive touch screen control is pictured in the bottom center of the figure. This control allows users to interactively explore different data sets and visualization schemes.